\documentclass[a4paper,showpacs,prl,twocolumn]{revtex4}
\usepackage{amsfonts}
\usepackage{amsmath}
\usepackage{amssymb}
\usepackage{graphicx}
\usepackage{verbatim}
\usepackage{color}
\usepackage{ulem}

\begin{document}

\title{Generic suppression of conductance quantization of interacting
electrons in graphene nanoribbons in a perpendicular magnetic field}
\author{A. A. Shylau}
\email{artsh@itn.liu.se}
\author{I. V. Zozoulenko}
\email{igozo@itn.liu.se}
\affiliation{Solid State Electronics, ITN, Link\"{o}ping University, 601 74, Norrk\"{o}%
ping, Sweden}
\author{H. Xu}
\author{T. Heinzel}
\affiliation{Condensed Matter Physics Laboratory, Heinrich-Heine-Universit\"at,
Universit\"atsstr.1, 40225 D\"usseldorf, Germany}

\begin{abstract}
The effects of electron interaction on the magnetoconductance of graphene
nanoribbons (GNRs) are studied within the Hartree approximation. We find
that a perpendicular magnetic field leads to a suppression instead of an
expected improvement of the quantization. This suppression is traced back to
interaction-induced modifications of the band structure leading to the
formation of compressible strips in the middle of GNRs. It is also shown
that the hard wall confinement combined with electron interaction generates
overlaps between forward and backward propagating states, which may
significantly enhance backscattering in realistic GNRs. The relation to
available experiments is discussed.
\end{abstract}

\date{\today }
 \maketitle




Conductance quantization in quantum point contacts (QPCs) and quantum wires
represents a hallmark of mesoscopic physics \cite{QPC,BvH}. At zero magnetic
field this effect can be understood within a noninteracting electron picture
as quantization of the transverse electron motion where, according to the
Landauer-Buttiker formalism, each propagating mode contributes with the
conductance quantum $G_{0}=2e^{2}/h$ to the total conductance \cite{QPC,BvH}%
. In a perpendicular magnetic field $B$ the propagating states acquire
qualitatively new features gradually transforming into edge states as $B$ is
increased \cite{BvH,edge states,Buttiker,Davies}. Since the left- and
right-propagating edge states get localized in transverse direction at
opposite wire edges in sufficiently strong magnetic fields, the coupling
between them can be exponentially small. This, in turn, leads to a strongly
suppressed backscattering and hence to a drastic improvement of the
conductance quantization\cite{BvH,edge states,Buttiker,Davies,vanWees}.
Taking electron interaction and screening in high magnetic fields into
account leads to new features such as formation of compressible and
incompressible strips\cite{Chklovskii}, which are essential for an
interpretation of various magnetotransport phenomena in conventional QPCs
and quantum wires defined in two-dimensional electron gases (2DEGs) \cite%
{Chklovskii,narrow channels}.

The isolation of graphene\cite{Novoselov} has immediately inspired the
search for conductance quantization in graphene nanoribbons (GNRs). However,
in all experiments reported so far conductance quantization at $B=0$ is
absent \cite{Han} or strongly suppressed \cite{Avouris}, which by now is
well understood and attributed to the effects of impurity scattering and/or
edge disorder \cite{disorder}. In analogy with conventional QPC structures
one would thus anticipate a drastic improvement of the conductance
quantization in GNRs in the edge state regime due to the expected
suppression of backscattering \cite{Buttiker}. Surprisingly enough, the
magnetoconductance measurements on GNRs reported so far show no evidence of
the expected improvement of the conductance quantization \cite%
{Molitor,Oostiga}. Even relatively large graphene strips ($\gtrsim 1\mu $m)
\cite{Marcus,FQHE} do not exhibit quantization plateaus at high magnetic
fields of high quality as routinely seen in corresponding conventional
heterostructures \cite{vanWees}.

In the present paper we study the magnetoconductance of GNRs taking electron
interaction on the Hartree level into account. Contrary to expectations
based on the conventional edge-state picture of noninteracting electrons
\cite{Buttiker} we find that application of a magnetic field leads to a
\textit{suppression} instead of expected improvement of the conductance
quantization. This behavior is related to a drastic modification of the GNR
band structure by electron interaction leading, in particular, to the
formation of compressible strips in the middle of the ribbon. These features
are generic in GNRs, but in contrast to most of the distinct properties of
graphene \cite{Castro Neto} they are not caused by the Dirac-like energy
dispersion but rather by the hard-wall confinement.

\begin{figure}[tbh]
\includegraphics[keepaspectratio,width=0.8\columnwidth]{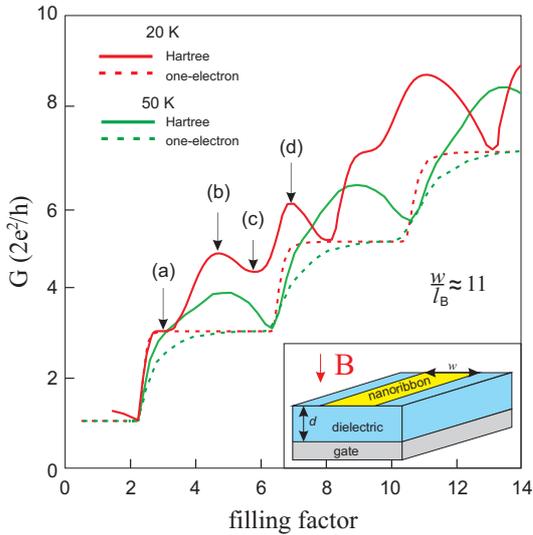}
\caption{(Color online) Conductance of the GNR as a function of filling
factor for interacting and noninteracting electrons at temperatures $T=20$ K
and 50 K in a magnetic field \textit{B}=30 T (corresponding to $l_{B}/w\approx11)$%
. The arrows indicate the filling factors for which the
corresponding band structures are shown in Fig.
(\protect\ref{fig:E(k)}) . Inset: sketch of the sample geometry. An
armchair GNR of width $w=50$ nm is located on top of an insulating
SiO$_{2}$ layer ($\protect\varepsilon _{r}=3.9$, thickness $d=300$
nm) and a gate electrode.} \label{fig:conductance}
\end{figure}

We consider a GNR attached to semi-infinitive leads acting as electron
reservoirs and subjected to a perpendicular magnetic field $B$, see inset to
Fig. \ref{fig:conductance}. The ribbon of width $w=50$ nm resides on top of
a SiO$_{2}$ insulating substrate ($\varepsilon _{r}=3.9)$ of thickness $d=300
$ nm, below which a metallic gate is located. The system is described by the
standard \textit{p}-orbital tight-binding Hamiltonian\cite%
{Wakabayashi,Castro Neto}
\begin{equation}
H=\sum_{\mathbf{r}}V_{H}(\mathbf{r})a_{\mathbf{r}}^{+}a_{\mathbf{r}}-\sum_{%
\mathbf{r},\Delta }t_{\mathbf{r},\mathbf{r}+\Delta }a_{\mathbf{r}}^{+}a_{%
\mathbf{r}+\Delta },  \label{Htb}
\end{equation}%
where the summation runs over all sites of the graphene lattice, $\Delta $
includes the nearest neighbors only, $t_{\mathbf{r},\mathbf{r}+\Delta
}=t_{0}\exp (i2\pi \phi _{\mathbf{r},\mathbf{r}+\Delta }/\phi _{0})$ with $%
t_{0}=2.77$ eV, $\phi _{0}=h/e$ being the magnetic flux quantum and $\phi _{%
\mathbf{r},\mathbf{r}+\Delta }=\int_{\mathbf{r}}^{\mathbf{r}+\Delta }\mathbf{%
A}\cdot d\mathbf{l}$ with $\mathbf{A}$ being the vector potential. We use
the Landau gauge, $\mathbf{A}=(-By,0)$. The interaction among the extra
charges of the density $n(\mathbf{r})$ is described within Hartree
approximation
\begin{equation}
V_{H}(\mathbf{r})=\frac{e^{2}}{4\pi \varepsilon _{0}\varepsilon _{r}}\sum_{%
\mathbf{r}^{^{\prime }}\neq \mathbf{r}}n(\mathbf{r}^{^{\prime }})\left(
\frac{1}{|\mathbf{r}-\mathbf{r}^{^{\prime }}|}-\frac{1}{\sqrt{|\mathbf{r}-%
\mathbf{r}^{^{\prime }}|^{2}+4d^{2}}}\right)   \label{Vr}
\end{equation}%
where the first term describes electron interaction within the ribbon while
the second term takes the presence of the metallic gate on the basis of the
image charge method into account. The band structure, the potential profile,
the charge density distribution are calculated self-consistently using the
Green's function technique (see Refs. \cite{Xu,Shylau} for details).
\begin{figure*}[tbh]
\includegraphics[keepaspectratio, width=2.0\columnwidth]{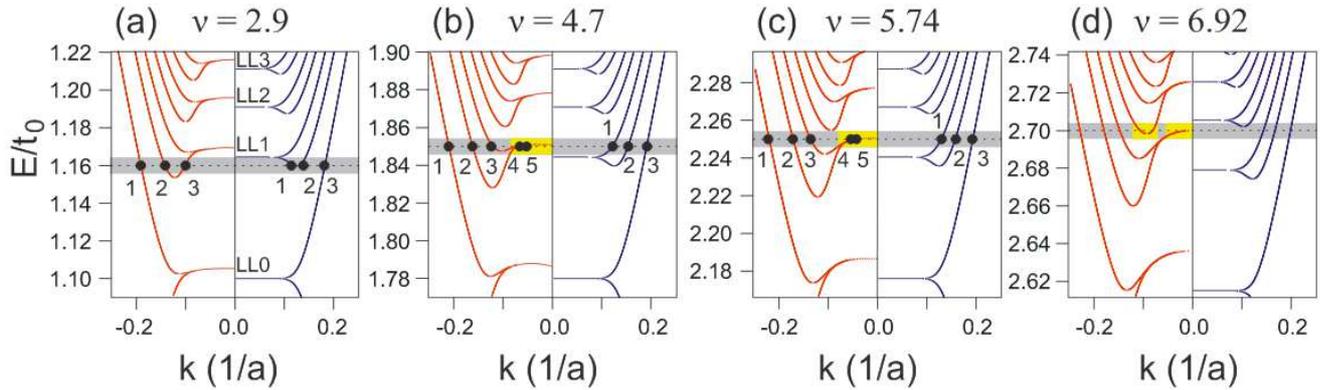}
\caption{(Color online) Evolution of the band structure of the GNR at
different filling factors corresponding to arrows (a) - (d) in Fig. \protect
\ref{fig:conductance}. Left and right parts of the panels correspond to the
interacting and non-interacting case, respectively. In order to align
noninteracting and Hartree bands the one-electron dispersions have been
shifted along the energy axis by the average Hartree energy. Gray fields
mark the energy window $[E_{F}-2\protect\pi k_{b}T,E_{F}+2\protect\pi k_{b}T]
$; yellow fields mark the compressible strips. The dotted line shows $E_{F}$%
. The black full circles mark the intersections of the Fermi level with the
dispersion curves, thereby identifying the propagating states at $E_{F}$. In
(a) the dispersionless states are marked according to the corresponding
Landau levels (LL) of the bulk graphene.}
\label{fig:E(k)}
\end{figure*}
The magnetoconductance through the nanoribbon in the linear response regime
is given by the Landauer formula
\begin{equation}
G(E_{F},B)=\frac{2e^{2}}{h}\int T(E,B)\left[ -\frac{\partial f_{FD}(E-E_{F})}{%
\partial E}\right] dE  \label{G}
\end{equation}%
where $f_{FD}(E-E_{F})$ is the Fermi-Dirac distribution function and $E_{F}$
denotes the Fermi energy. For an ideal system (without scattering), the
total transmission coefficient $T(E,B)$ is equal to the number of
propagating states, $T(E,B)=N_{prop}$, such that the conductance is simply
proportional to $N_{prop}$ weighted by $\frac{\partial f_{FD}}{\partial E}$
which is different from zero in an energy window $\sim 4\pi k_{B}T$.

Figure \ref{fig:conductance} shows the conductance of the ideal nanoribbon
for a representative magnetic field $B=30$ T as a function of the filling
factor $\nu =\left\langle n\right\rangle \phi _{0}/B$ for two representative
temperatures, with $\left\langle n\right\rangle $ being the electron density
averaged across the ribbon. Here, $\nu $ is tuned by varying the gate
voltage $V_{g}$ which is applied vs. the grounded nanoribbon and thus tunes
the electron density. The ratio of GNR width to magnetic length, $%
w/l_{B}\approx 11,$ is chosen in accordance with typical experiments \cite%
{Molitor, Oostiga}. It is important to emphasize that the obtained results
remain practically unchanged when the system is scaled by, e.g. increasing $%
w $ while simultaneously reducing $B$ such that the ratio $w/l_{B}$
remains constant. In order to highlight the role of electron
interaction, we compare our self-consistent calculations with a
noninteracting picture. The calculated conductance shows a striking
difference between the interacting and noninteracting cases. First
of all, at a given filling factor, the conductance of the
interacting system is always larger than that one of the
corresponding noninteracting system. Second, the perfect
quantization steps calculated for the noninteracting picture are
destroyed as the interaction is turned on, and the conductance
develops pronounced bump-like features. Note that the elevated
temperature smears the conductance bumps to some extent, but they
still dominate the conductance even at $T=50K.$ We also note that we
performed the magnetotransport calculations for a high-k material
($\varepsilon _{r}=47$) and a gate closeby, $d=5$ nm when the
electron interaction is strongly screened (not shown here). We find
that even in this case the bumps are suppressed but still clearly
dominate the conductance.

We proceed by interpreting the suppression of conductance plateaux in terms
of interaction-induced modifications of the energy dispersion. The evolution
of the band structure as a function of $\nu $ in the interval covering a
representative bump, $2.9\lesssim \nu \lesssim 5.7$ (corresponding to arrows
(a)-(c) in Fig. \ref{fig:conductance}) is presented in Fig. \ref{fig:E(k)}
both for interacting and non-interacting cases. The dispersion relation for
non-interacting electrons shows flat regions corresponding to the Landau
levels in bulk graphene\cite{McClure,Castro Neto} and dispersiveness states
close to the GNR's boundaries representing familiar edge states \cite{Davies}%
. Note that the position and the number of propagating states at a
given energy are determined by the intersection of the Fermi energy
level with the corresponding subbands.

For non-interacting electrons changing the gate voltage results in a
shift of the Fermi level but does not modify the subband structure.
Qualitatively new features arise when the electron interaction is
taken into account. One of the most distinct features is that the
dispersionless state in the center of the GNR (corresponding to the
1st Landau level (LL)) gets pinned to the Fermi energy thus forming
a compressible strip. These strips are marked in yellow in Fig.
\ref{fig:E(k)} (b)-(d); following Suzuki and Ando \cite{Suzuki} we
define a compressible strip as a region where the dispersion lies
within the energy window $|E-E_{F}|<2\pi k_{B}T$. The compressible
strips form because
in the above energy window the states are partially filled (i.e. $0<f_{FD}<1$%
) and hence the system has a metallic character. Due to the metallic
behavior, the electron density can easily be redistributed in order to
effectively screen the external potential \cite{Chklovskii}. The
compressible strips can form only if the confining potential is sufficiently
smooth \cite{Chklovskii}. The GNRs have a hard-wall confinement and hence
the compressible strips can form only in the center but not for the edge
states. Note that the existence of compressible strips in graphene has been
recently demonstrated by Silvestrov \textit{et al.}\cite{Efetov}.

Because of the pinning of the LL to the Fermi energy, changing of
the filling factor leads to a significant distortion of the
dispersion curves. For a given $B$, the larger the gate voltage (and
therefore $\nu $) is, the stronger the bands are distorted in
comparison to the noninteracting picture (c.f. (a)-(d) in Fig.
\ref{fig:E(k)} ). This distortion eventually leads to the bumps in
the conductance. Indeed, according to Eq. (\ref{G}) the conductance
is given by the number of propagating states averaged in the energy
window $|E-E_{F}|<2k_{B}T$. For noninteracting electrons the
dispersion relation is not changed as $\nu $ is varied and the
number of propagating states remains always the same, $N_{prop}=3$
(see right panels in Fig. \ref{fig:E(k)} (a)-(c)). This, according
to Eq. (\ref{G}), leads to a conductance plateau $G=3G_{0}$. In
contrast, for interacting electrons the dispersion relation gets
distorted and there is always an energy interval in the window
$|E-E_{F}|<2k_{B}T$ where the number of propagating states exceeds
that for the non-interacting case. This is illustrated in the left
panels in Fig. \ref{fig:E(k)} (b)-(c) for $E=E_{F}$ where
$N_{prop}=5$. As a result, the conductance exceeds its
noninteracting value of $3G_{0}$ exhibiting the pronounced bumps.

With further increase of $\nu $, the compressible strips pinned to
the Fermi level form not only in the center of the strip, but
further away from the center (as illustrated in Fig. \ref{fig:E(k)}
(d)). Note that the second compressible strip in Fig. \ref{fig:E(k)}
(d) leads to the formation of a bump in the conductance in the
region $6\lesssim \nu \lesssim 9$.

\begin{figure}[tb]
\includegraphics[keepaspectratio,width=\columnwidth]{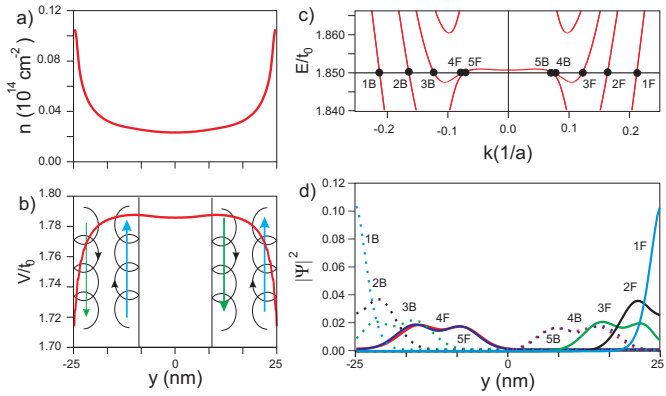}
\caption{(Color online) (a) Electron concentration $n(y)$, (b)
self-consistent potential $V_H$ across the GNR, and (c) the band
structure at $\protect\nu =4.7$. (d) The square modulus of the wave
functions at $E_F$ of forward (F) and backward (B) propagating
states (solid and dashed lines correspondingly) marked in (c). For
the sake of clarity the electron densities, potential and the wave
functions are averaged over two adjacent slices and 3 adjacent sites
of the same slice). Inset in (b) illustrates classical skipping
orbits.} \label{fig:states}
\end{figure}

Let us now discuss in detail a structure of propagating states of
the interacting electrons in GNRs. Figures \ref{fig:states} (a), (b)
show the electron density and the confining potential for a
representative filling factor $\nu =4.7$ [(b)-arrow on Fig.
(\ref{fig:conductance})]. The distribution of charge density is
highly nonuniform showing charge accumulation at the boundaries
\cite{Efetov,Shylau}. There are two types of states, which have a
different microscopic character. The first type [1,2,3 states in
(a)-(c)] corresponds to edge states propagating near the boundaries
and have the same structure for interacting and noninteracting
cases. The second type [states 4 and 5] corresponds to the states
which form compressible strips in the center of the ribbon as
discussed above. The most prominent feature of these states is that
their direction of propagation is opposite to that one of the edge
states residing in the same half of the GNR. This is in contract to
the noninteracting picture, where due to the presence of a magnetic
field, forward and backward propagating states are localized at
different boundaries by Lorentz forces. This unusual behavior can be
interpreted in terms of a semiclassical analogue. The electrons
scattered at the boundaries are described by skipping orbits.
Besides the hard-wall potential walls provided by nanoribbon's
edges, there are two additional walls originating from the
self-consistent potential which, together with the outer walls of
the GNR, form triangular quantum wells at the ribbon's edges [Fig.
\ref{fig:states} (b)]. Electrons which strike the left side of the
right triangular quantum well propagate in the same direction as the
electrons that strike the left edge of the nanoribbon as
schematically illustrated in Fig. \ref{fig:states} (b).

This feature of propagating states in high magnetic field makes GNRs much
more sensitive to the effect of the disorder in comparison to conventional
split-gate structures defined in 2DEG. Indeed, for interacting electrons in
GNRs the overlap between the backward (4B, 5B) and forward (1F-3F)
propagating states is significant. In realistic GNRs with disorder this
would result in a strong enhancement of backscattering, which, in turn, can
lead to a further distortion of the conductance (in addition to bumps that
are present even in ideal GNRs without disorder).

Note that the features of the band structure and character of propagating
states in GNRs discussed above are not caused by the Dirac-like energy
dispersion but rather by the hard-wall confinement at the boundaries. These
features of the GNRs resemble corresponding features of cleaved-edge
overgrown (CEO) quantum wires \cite{Ihnatsenka} that also have a hard-wall
confinement. We therefore expect that magnetoconductance of CEO also should
exhibit suppressed quantization of high field (Note that we were not able to
find any reports on magnetoconductance measurements in CEO wires at high
magnetic field).

We continue by relating our results to the available experimental data. We
are not aware of any studies reporting a drastic improvement of the
conductance quantization in GNRs by perpendicular magnetic field. The
observed conductance in narrow GNRs exhibit irregular \cite{Oostiga} or
bump-like features \cite{Molitor}, and the wider structures show pronounced
bumps superimposed on conductance plateaus \cite{Marcus,FQHE}. Even though
this is consistent with our findings, this can hardly be regarded as a
definite experimental validation of our predictions. We thus hope that our
work will motivate systematic studies of the magnetoconductance that will
shed new light on properties of interacting electrons in confined graphene
systems.

In conclusion, we have shown that applying a perpendicular magnetic
field to a GNR containing an interaction electron gas leads to a
\textit{suppression} instead of expected improvement of conductance
quantization. This surprising behavior is related to the
modification of the band structure of the GNR due to the electron
interaction leading, in particular, to the formation of compressible
strips in the middle of the ribbon and existence of
counter-propagating states in the same half of the GNR.

A.A.S. and I. V. Z acknowledge a support of the Swedish Research Council
(VR).

\end{document}